\documentclass[12pt]{article}
\usepackage{amsmath}
\usepackage{amssymb}
\usepackage{mathtext}
\usepackage{graphicx}

\begin{document}

\section*{ON JUSTIFICATION OF GIBBS DISTRIBUTION\footnotemark}
\footnotetext{REGULAR AND CHAOTIC DYNAMICS, V. 7, No. 1, 2002

{\it Received January 10, 2001

AMS MSC 37H10, 70F45}}

\begin{centering}
V.\,V.\,KOZLOV\\
Department of Mechanics and Mathematics\\
Moscow State University\\
Vorob'ievy gory, 119899 Moscow, Russia\\
\end{centering}

\begin{abstract}
The paper develop a new approach to the justification of Gibbs canonical distribution
for Hamiltonian systems with finite number of degrees of freedom. It uses the condition of
nonintegrability of the ensemble of weak interacting Hamiltonian systems.
\end{abstract}

\paragraph{1. Gibbs distribution.}

We consider the probability distribution in the phase space of Hamiltonian
system with the density
\begin{equation}
\rho=ce^{-\frac H{k\tau}},
\end{equation}
where $H$ is a Hamiltonian, $\tau$ is an absolute temperature, $k$ is the
Boltzmann constant, $c$ is a normalized factor. It plays the key role in
the equilibrium statistical mechanics. Gibbs show in~[1] that the
averaging with respect to probability measure with density~(1) give rise
to the fundamental relations of equilibrium thermodynamics.

To deduce the canonical Gibbs distribution one usually consider the
ensemble of Hamiltonian systems with Hamiltonian function of the following
form
\begin{equation}
\mathcal{H}=\mathcal{H}_0(P,\,Q)+\varepsilon\mathcal{H}_1(P,\,Q),
\end{equation}
where\vspace{-6mm}
\begin{gather}
\mathcal{H}_0=\sum\limits_{s=1}^n H_0(p^{(s)},\,q^{(s)}),\notag\\
p^{(s)}=(p^{(s)}_1,\,\ldots,\,p^{(s)}_m),\qquad
q^{(s)}=(q^{(s)}_1,\,\ldots,\,q^{(s)}_m).
\end{gather}
Thus at $\varepsilon=0$ the system with Hamiltonian~(2) is decomposed
on~$n$ identical systems with~$m$ degrees of freedom and
Hamiltonian~$H_0$. The canonical variables~$P$,~$Q$ are the
momenta~$p^{(1)},\,\ldots,\,p^{(n)}$ and
coordinates~$q^{(1)},\,\ldots,\,q^{(n)}$ of separate subsystems. The
perturbing function~$\mathcal{H}_1$ is the energy of interaction of
$n$~subsystems; it usually depends on their coordinates~$Q$. Small
parameter~$\varepsilon$ is the characteristic of intensity of subsystems'
interaction.

We consider the case, when the Hamiltonian~$\mathcal{H}$ is sufficiently
smooth with respect to variables~$P$,~$Q$. However, in application we
often see cases with singular interaction. The classical example is the
Boltzmann-Gibbs gas, the assembly of rigid balls in cube that elastically
collide with each other (see [1,\,2,\,3]).

The traditional approach to the deduction of Gibbs distribution suggested
by Fowler and Darwin ([4], the rigorous exposition see in~[5,\,6])
essentially uses the ergodic hypothesis: for all small~$\varepsilon>0$ the
Hamiltonian system with Hamiltonian~(2) is ergodic on fixed energy
manifolds~$\mathcal{H}=const$. With some additional conditions some
systems with Hamiltonian~$H_0$ are distributed in accordance with
formula~(1) as~$\varepsilon\to0$ and~$n\to\infty$.

But the proof of ergodic hypothesis for specific Hamiltonian systems is
usually pretty difficult problem. Moreover the ergodic hypothesis often
contradicts with results of KAM theory. In particular if the Hamiltonian
system with Hamiltonian~$H_0$ is completely integrable and the energy
surfaces~$H_0=const$ are compact, then the ergodic property is not
present with certainty.

In view of this remark we can set the following interesting problem: prove
that in analytical (or even in infinitely differentiable) case if the
energy surfaces~$H_0=const$ are compact, then the Hamiltonian system with
Hamiltonian~(2) never satisfy ergodic hypothesis.

We can try to modify the Fowler--Darwin method assuming that
$\varepsilon\ne0$ and~$n$ depend on~$\varepsilon$ in such way
that~$n(\varepsilon)\to\infty$ as~$\varepsilon\to0$. We can assume that
for some functions~$\varepsilon\mapsto n(\varepsilon)$ the system with
Hamiltonian~(2) is ergodic on the surfaces~$\mathcal{H}=const$ as a
result of huge number of its degrees of freedom. This somewhat weakened
version of ergodic hypothesis is closely related to the unsolved problem
of estimation of small parameter in KAM theory, when the ``last'' Kolmogorov
torus disappears. The weaker conjecture on transitivity of system with
Hamiltonian~(2) on energy surfaces~$\mathcal{H}=const$ for large values
of~$n$ and small~$\varepsilon$ is not proved yet. If this problem has the
positive answer, then we can assert the presence of diffusion in
Hamiltonian systems with many degrees of freedom (see [7--9]).

\paragraph{2. Probability density as an integral of Hamiltonian equations.}

A different approach to the deduction of canonical Gibbs distribution was
proposed in the paper [10]. This approach is based on the fact that the
stationary density of probability distribution is the integral of
Hamiltonian differential equations uniquely defined in the whole phase
space. In this case the number of interacting subsystem~$n\ge2$ is fixed.

More exactly in [10] we consider the case, when the subsystems have one
degree of freedom:~${m=1}$. Under some natural condition we can proceed to
the angle--action variables in each subsystem, and using the well-known
Poincar\'{e} method we can obtain the constructive condition of
nonexistence of new integrals (see [11,\,12]). The results of paper~[10]
can be easily converted to the more general case, when~$H_0$ is a
Hamiltonian of completely integrable system. We are going to find out the
sufficient conditions (constructive if possible) of nonintegrability of
systems~(2)--(3).

We study the conditions of existence of an integral
$\mathcal{F}(P,\,Q,\,\varepsilon)$ of the canonical differential equations
\begin{equation}
\dot{P}=-\frac{\partial\mathcal{H}}{Q},\qquad
\dot{Q}=\frac{\partial\mathcal{H}}{P}
\end{equation}
with a Hamiltonian $\mathcal{H}$ of the form~(2)--(3). We emphasize
that the integral~$\mathcal{F}$ depends on the parameter~$\varepsilon$.
Poincar\'{e} considered the analytical case; in particular we can
construct the integral~$\mathcal{F}$ as a power series with respect
to~$\varepsilon$. We suppose that $\mathcal{F}$ is a function of
class~$C^2$ with respect to all the variables~$P$,~$Q$ and~$\varepsilon$.
Hence we can suppose
\begin{equation}
\mathcal{F}=\mathcal{F}_0(P,\,Q)
+\varepsilon\mathcal{F}_1(P,\,Q)+o(\varepsilon),
\end{equation}
where $\mathcal{F}_0$ and~$\mathcal{F}_1$ are functions of class~$C^2$
and~$C^1$ with respect to~$P$ and~$Q$ correspondingly. Probably we can
weaken the requirements to the class of smoothness of
integral~$\mathcal{F}$, and the following arguments still will be correct.
But this is a separate problem and we are not going to discuss it here.

\paragraph{3. Non-perturbed problem.}
Suppose $\varepsilon=0$. Then we have a system of~$n$ independent
subsystems. It is strongly nonergodic: at~$\varepsilon=0$ system of
differential equations~(4) has~$n$ independent first integrals
\begin{equation}
H_s=H_0(p^{(s)},\,q^{(s)}),\qquad 1\le s\le n.
\end{equation}
It is clear that the function $\mathcal{F}_0$ from expansion~(5) is a
first integral of this unjointed system. Let's show that under the
specific conditions the function~$\mathcal{F}_0$ depends only
on~$H_1,\,\ldots,\,H_n$. In particular these conditions will imply that
any separate subsystem with~$m$ degrees of freedom does not nave first
integrals independent on the integral of energy. The ideas of our
arguments follows Poincar\'{e} method~[11].

Let $M$ be a phase space of Hamiltonian system with Hamiltonian
function~(6). Certainly for all~$s$ these spaces are identical. A phase
space of new system is the direct product
$$
\mathcal{M}=M\times\ldots\times M,\qquad \dim\mathcal{M}=2nm.
$$
Let $h_s$ be a value of total energy of system with number $s$, and
$$
\sum(h_s)=\{p^{(s)},\,q^{(s)}:H_s(p^{(s)},\,q^{(s)})=h_s\}
$$
is the a energy surface. If the value of~$h_s$ is uncritical, then $\sum$
is a smooth $2m-1$-dimensional manifold. At fixed values
of~$h=(h_1,\,\ldots,\,h_n)$ and~$\varepsilon=0$ unjoined Hamiltonian
system~(4) is reduced to the direct product of dynamical systems defined
on
$$
S(h)=\sum(h_1)\times\ldots\times\sum(h_n)\subset\mathcal{M}.
$$

The ergodic property of system with Hamiltonian~$H_s$ on~$\sum(h_s)$ does
not necessarily imply the constancy of integral~$\mathcal{F}_0$ on the
invariant set~$S(h)$. Consider the following simple example

{\it EXAMPLE.}
Suppose the following dynamical system
\begin{equation}
\dot{x}_i=\omega_i,\qquad \dot{y}_j=\omega_j;\qquad
i,\,j=1,\,\ldots,\,k
\end{equation}
with constant incommensurable systems
$\omega=(\omega_1,\,\ldots,\,\omega_k)$ is defined on direct product of
$k$-dimensional tori $\mathbb{T}^k\{x\bmod
2\pi\}\times\mathbb{T}^k\{y\bmod 2\pi\}$. According to the Weyl theorem,
each separate subsystem is ergodic on~$\mathbb{T}^k$. But equations~(7)
have single-valued nonconstant integrals $\sin(x_i-y_i)$\, ($1\le i\le
k$).

{\bf Remark.} {\it
However, if Hamiltonian systems are weakly mixing (mixing) systems
on~$\sum(h_s)$, then their direct product also has weakly mixing (mixing)
property on~$S(h)$. In particular in these cases the
function~$\mathcal{F}_0$ is constant on any connected component of
manifold~$S(h)$.}

Let $\mathbb{T}^1$ be a nondegenerate periodic trajectory of system with
number $s$ with energy~$h_s$, $T_s$ its period, and~$\omega_s=2\pi/T_s$
its frequency. According to Floquet--Lyapunov theorem, in a neighborhood
of this trajectory on~$\sum(h_s)$ we can express the
coordinates~$\varphi_s\bmod 2\pi$,\,$z_1^{(s)},\,\ldots,\,z_{2m-2}^{(s)}$,
so that in the new variables the motion equation obtain the following
form:
\begin{equation}
\dot{\varphi}_s=\omega_s+f_s(\varphi_s,\,z^{(s)}),\qquad
\dot{z}^{(s)}=\Omega_sz^{(s)}+g_s(\varphi_s,\,z^{(s)}).
\end{equation}
Here $f_s=O(|z^{(s)}|)$, $g_s=o(|z^{(s)}|)$, and~constant square
matrix~$\Omega_s$ of order~$2m-2$ is nondegenerate. Assuming in~(8)
$z^{(s)}=0$ we obtain the equation of periodic trajectory:
\begin{equation}
\dot{\varphi}_s=\omega_s\qquad (1\le s\le n).
\end{equation}

According to the assumption on non degeneracy of periodic
trajectory~$\mathbb{T}^1$, nondegenerate periodic trajectories with
similar period are situated on close energy surfaces~$\sum(h_s)$; periods
and frequencies continuously depend on energy~$h_s$.

It is clear that the direct product
$\mathbb{T}^1\times\ldots\times\mathbb{T}^1=\mathbb{T}^n$ is
$n-$dimensional invariant torus of canonical system of differential
equations~(4) at~$\varepsilon=0$ situated on~$S(h)$. In the neighborhood
of this torus the equation of motion have form (8). Hence, such torus is
reducible and nondegenerate (see, for example, [12]). On the torus the
equation are reduced to a conditionally-periodic form (9). As usually,
we call an invariant torus nonresonance if the frequencies
$\omega_1,\,\ldots,\,\omega_n$ are independent on the ring of integers. In
the further analysis the following condition is essential

\hangindent=40pt\hangafter=1 A) For almost all admissible values of
$h\in\mathbb{R}^n$ nonresonance tori are everywhere dense on the manifold
$S(h)$.

{\it EXAMPLE.}
Let separate subsystems describe the inertial motion on the manifold
$\mathcal{N}$ of negative curvature. Energy $h$ is non-negative. All
periodic trajectories with positive energy are hyperbolical; hence they
are not degenerate. Periodic trajectories are the motions on closed
geodetics on $\mathcal{N}$ with different speed. If $l$ is a length of
closed geodetic, then the period is equal to $\frac{l}{\sqrt{2h}}$. Hence
the frequency $\omega$ is defined by the formula\vspace{-3mm}
$$
2\pi\frac{\sqrt{2h}}{l}.
$$
Since the lengths of $n$ geodetics $l_1,\,\ldots,\,l_n$ are fixed, then
for almost all positive values of energy $h_1,\,\ldots,\,h_n$ the
frequencies $\omega_1,\,\ldots,\,\omega_n$ are incommensurable.
It is possible to show (and this is a separate problem) that in a
considered situation condition A is fulfilled.

{\bf Proposition 1.} {\it
If condition A is fulfilled, then for all $h\in\mathbb{R}^n$ the
function $\mathcal{F}_0$ is constant on any connected component of $S(h)$.}

{\it Proof.} Since $\mathcal{F}_0$ is an integral of system of equations (4)
at $\varepsilon=0$, then (according to the Kronecker theorem)
$\mathcal{F}_0$ is constant on any nonresonance torus $\mathbb{T}^n$.
Since this torus is reducible and nondegenerate, then $d\mathcal{F}_0=0$
in points $\mathbb{T}$ (see [12], ch. IV). According to condition A, for
almost all values of $h\in\mathbb{R}^n$ nonresonance tori are everywhere
dense on $S(h)$. Hence, $d\mathcal{F}_0=0$ on such manifolds $S(h)$.
Therefore $\mathcal{F}_0$ is constant on their connected components. For
other values of $h$ the conclusion of proposition 1 follows by continuity.

{\bf Remark.} {\it
The proof shows that in condition A instead of ``for almost all admissible
values of $h\in\mathbb{R}^n$'' we can say ``for everywhere dense set of
values of $h\in\mathbb{R}^n$''. However, such weakening of condition A
practically does not give anything new.}

In further analysis we will use the proposition on everywhere density of
the set of maximum resonance tori (when all frequencies are rationally
expressed through one frequency). This condition together with condition A
produces ``an alternation'' of resonance and nonresonance invariant tori
and replaces the condition of nondegeneracy of non-perturbed completely
integrable system in the Poincar\'{e} theory.

\paragraph {4. Energy surfaces.}

Let us consider a case, when function the $H_0:M\to\mathbb{R}$ has a
finite number of critical values $a_1<a_2<\ldots<a_r$, and $a_1=\min H_0$.
Such situation is common in applications. When the total energy $h_0$
passes through the critical value, the continuous dependence of energy
surface $\sum(h_0)$ on $h_0$ is lost. In that moment its topology
generally changes.

%\fig<bb=0 0 75.1mm 55.5mm>{graphic2.eps}

In fig. 1 we present the plot of Hamiltonian $H_0$ with three critical
values. The points $a_1$ and $a_3$ are stationary values of $H_0 $, and
the critical point $a_2$ is not a stationary value. The presence of the
nonstationary critical points is the characteristic property of potentials
describing gravitational or Coulomb interaction.

Let's denote as $K_{i_1 i_2\,\ldots\,i_n}$ an open parallelepiped in
$\mathbb{R}^n=\{h_1,\,\ldots,\,h_n\}$. This parallelepiped is a
direct product of the intervals
\begin{equation}
a_{i_1}<h_1<a_{i_1+1},\,\ldots,\,a_{i_n}<h_n<a_{i_n+1}.
\end{equation}
If number $i_s+1$ is larger than $r$, then we replace $a_{i_s+1}$ with a
symbol $\infty$. In fig. 2 these domains are shown for $n=2$ and $r=3$.
Each point $h\in K_i$,\,$i=(i_1,\,\ldots,\,i_n)$ corresponds to a smooth
regular manifold, which may consist of several connected parts. The
quantity of connected components of $S(h)$ does not depend on a point
$h\in K_i$; we denote this number by symbol $\varkappa_i$.

%\fig<bb=0 0 106.0mm 67.1mm>{graphic1.eps}

Let us introduce in the phase space $\mathcal{M}$ open areas
$\mathcal{K}_i$ defined by the inequalities similar to (10):
$$
a_{i_1}<H_1(p^{(1)},\,q^{(1)})<a_{i_1+1},\,\ldots,\,a_{i_n}<
H_n(p^{(n)},\,q^{(n)})<a_{i_n+1}.
$$

It is clear that the closure of these domains in the whole covers all
$\mathcal{M}$. Also each $\mathcal{K}_i$ has exactly $\varkappa_i$
connected components.

{\bf Proposition 2.} {\it
For any connected component of domain $\mathcal{K}_i$ there exists the
continuously differentiable function}
$$
f_i:K_i\to\mathbb{R},
$$
{\it such that the following equality is fulfilled in this domain}
\begin{equation}
\mathcal{F}_0=f_i(H_1,\,H_2,\,\ldots,\,H_n).
\end{equation}

{\bf Remark.} {\it
Actually function $f_i$ belongs to the class of smoothness $C^2$. However
it is not essential for the further analysis.}

{\it Proof.} It is clear that the domain $\mathcal{K}_i$ is foliated to the
regular surfaces $S$. The function $\mathcal{F}_0$ is constant on these
surfaces (more exactly on their connected components) (the proposition 1).
Hence, on any connected component $\mathcal{K}_i$ the function
$\mathcal{F}_0$ has natural representation (11). By definition in any
point $\mathcal{K}_i$ functions $H_1,\,\ldots,\,H_n$ are independent.
Therefore we can introduce locally new coordinates so that
$H_1,\,\ldots,\,H_n$ will appear as $n$ of new variables. The transition
to such coordinates is carried out with certainty with the help of
continuously differentiable reversible transformation. In new variables
the function $\mathcal{F}_0$ is continuously differentiable and depends
only on $n$ variables $H_1,\,\ldots,\,H_n $. The proposition is proved.

\paragraph{5. Resonances.}
Let us consider again an invariant torus $\mathbb{T}$ of non-perturbed
system. The equations of motion on the torus are reduced to form (9).
This torus we call completely resonance if there exist $n-1$ linearly
independent integer vectors
$$
u=(u_1,\,\ldots,\,u_n),\quad v=(v_1,\,\ldots,\,v_n),\,\ldots, \quad
w=(w_1,\,\ldots,\,w_n),
$$
such that
\begin{equation}
(u,\,\omega)=u_1\omega_1+\ldots+u_n\omega_n=0,\quad
(u,\,\omega)=0,\,\ldots, \quad (w,\,\omega)=0.
\end{equation}
In other words all frequencies $\omega_s$ are rationally expressed through
one of them. This is equivalent to the proposition that all solutions of
differential equations (9) on the torus $\mathbb{T}^n$ are periodic with
the same period.

Let $\Phi$ be a restriction of perturbing function
$\mathcal{H}:M\to\mathbb{R}$ on the invariant torus $\mathbb{T}^n$. It is
clear that $\Phi$ is $2\pi-$periodic function of
$(\varphi_1,\,\ldots,\,\varphi_n)=\varphi$. We associate it with multiple
Fourier series
\begin{equation}
\Phi=\sum_{k\in\mathbb{Z}^n}\varphi_k\exp i(k,\,\varphi).
\end{equation}

A nonresonance torus we call the Poincar\'{e} torus if the factors of
Fourier decomposition (13) with numbers $u,\,v,\,\ldots,\,w$ are nonzero.
Since the Poincar\'{e} tori consist of the separate closed trajectories,
they have no ``rigidity'' property and collapse after the addition of
perturbation. We can not exclude the possibility that the family of
degenerate periodic trajectories, components of the Poincar\'{e} torus, at
perturbation give rise to the finite number of nondegenerate periodic
trajectories with close transition.

Let us introduce finally the Poincar\'{e} set
$\mathbb{P}\subset\mathbb{R}^n$. It is a set of points from
$\mathbb{R}^n=\{h_1,\,\ldots,\,h_n\}$, which are the images of the
Poincar\'{e} tori under the ``energy'' mapping
$\mathcal{M}\to\mathbb{R}^n$: a point with coordinates
$$
(P,\,Q)=(p^{(1)},\,\ldots,\,p^{(n)},\:q^{(n)},\,\ldots,\,q^{(n)})
$$
passes to a point with coordinates
$$
h_1=H_1(p^{(1)},\,q^{(1)}),\,\ldots, \quad h_n=H_n(p^{(n)},\,q^{(n)}).
$$

{\bf Remark.} {\it
In the Poincar\'{e}} {\it theory [11,\,12] the usually supposition is that the
non-perturbed system with the Hamiltonian $\mathcal{H}_0$ is completely
integrable and nondegenerate. Therefore the set of its first integrals is
a set of action variables ``numerating'' the invariant tori. The
Poincar\'{e}} {\it set (introduced in [12]) is defined here as a set of points
in space of action variables corresponding to the completely resonance
tori collapsing after the addition of perturbation. In our case functions
$H_1,\,\ldots,\,H_n$ are the set of integrals of the non-perturbed problem
and the Poincar\'{e}} {\it set is a set of points in space of values of these
integrals.}

{\bf Proposition 3.} {\it
In points of the Poincar\'{e} set $\mathbb{P}$} {\it functions}
$$
\mathcal{H}_0=\sum_{s=1}^n\qquad \text{\it and}\qquad
\mathcal{F}_0=f(H_1,\,\ldots,\,H_n)
$$
{\it are dependent.}

{\it Proof.} Let $\{\,\}$ be a Poisson bracket connected with symplectic
structure on $\mathcal{M}$. Since function $\mathcal{F}$ is the first
integral of initial system (4), then $\{\mathcal{H},\,\mathcal{F}\}=0$
for all values of $\varepsilon$. Since
$\{\mathcal{H}_0,\,\mathcal{F}_0\}=0$, then
\begin{equation}
\lim\limits_{\varepsilon\to 0}
\frac{\{\mathcal{H},\,\mathcal{F}\}}{\varepsilon}=0.
\end{equation}
Using decomposition (2) and (5) we receive from (14) the equality
\begin{equation}
\{\mathcal{H}_0,\,\mathcal{F}_1\}=\{\mathcal{F}_0,\,\mathcal{H}_1\}.
\end{equation}
Well known that $\{\mathcal{F}_0,\,\mathcal{H}_1\}$ is a derivative of
$\mathcal{F}_1$ by virtue of the initial system of differential equations
with Hamiltonian $\mathcal{H}_0$. The Poisson bracket
$\{\mathcal{F}_0,\,\mathcal{H}_1\}$ has similar sense. Besides by formula
(11) we obtain
\begin{equation}
\{\mathcal{F}_0,\,\mathcal{H}_1\}=\sum_{s=1}^n \frac{\partial f}{\partial
H_s}\{H_s,\,\mathcal{H}_1\}.
\end{equation}

Now let us restrict equality (15) on the invariant torus $\mathbb{T}^n$.
It is clear that $\{\mathcal{H}_0,\,\mathcal{F}_1\}$ is equal to the
derivative of restriction of function $\mathcal{F}_1$ on $\mathbb{T}^n$ by
virtue of system of differential equations (9). Let
$\Psi:\mathbb{T}^n\to\mathbb{R}$ be a restriction of $\mathcal{F}_1$ on
$\mathbb{T}^n$ and
\begin{equation}
\Psi=\sum_{k\in\mathbb{Z}^n}\psi_k\exp i(k,\,\varphi)
\end{equation}
its Fourier series. In points $\varphi\in\mathbb{Z}^n$ the left part of
(5.4) becomes
\begin{equation}
\sum_{k\in\mathbb{Z}^n} i(k,\,\omega)\psi_k\exp i(k,\,\varphi).
\end{equation}

In view of formulas (13) and (16) the right part of relation (15) in
points of the invariant torus is equal to
\begin{equation}
\sum_{k\in\mathbb{Z}^n}i\left(\frac{\partial f}{\partial H_1}k_1\omega_1
+\ldots+\frac{\partial f}{\partial H_n}k_n\omega_n\right) \varphi_k\exp
i(k,\,\varphi).
\end{equation}

Comparing (18) and (19) we obtain a chain of equalities
\begin{equation}
(k,\,\omega)\psi_k=(\xi k_1\omega_1+\ldots+\xi_n k_n\omega_n)
\psi_k,\qquad k\in\mathbb{Z}^n.
\end{equation}

Now let $\mathbb{T}^n$ be the Poincar\'{e} torus. Setting $k$ equal to
$u,\,v,\,\ldots,\,w$ we obtain that $\omega$ as a vector of
$n-$dimensional space is orthogonal to the hyperplane $\Pi$ generated by
linearly independent vectors $u,\,v,\,\ldots,\,w$. Since
$(u,\,\omega)=\ldots=(w,\,\omega)=0 $, and
$\varphi_u\ne0,\,\ldots,\;\varphi_w\ne0$, then we obtain from
(20) $n-1$ linear relations:
$$
\begin{matrix}
u_1(\xi_1\omega_1)+\ldots+u_n(\xi_n\omega_n)=0,\\
\hdotsfor{1}\\
w_1(\xi_1\omega_1)+\ldots+w_n(\xi_n\omega_n)=0.
\end{matrix}
$$
Hence the vector with components $\xi_1\omega_1,\,\ldots,\,\xi_n\omega_n$
is orthogonal to the hyperplane $\Pi$ and consequently it is collinear to
the vector $\omega=(\omega_1,\,\ldots,\,\omega_n)$. Hence
$\xi_1=\ldots=\xi_n$.

Thus in points of the Poincar\'{e} set the derivatives
$\frac{\partial f}{\partial H_1},\,\ldots,\,\frac{\partial f}{\partial
H_n}$ are equal to each other. It means obviously the dependence of
functions $\mathcal{H}_0$ and $\mathcal{F}_0$.

{\bf Remark.} {\it
Since functions $\Phi$ and $\Psi$ are assumed to be only once continuously
differentiable, then the Fourier series (13), (17)-(19) may diverge. In
this case relations (20) can be deduced in another way. For this purpose
we multiply the derivative with respect to time}
$$
\dot{\Phi}=\frac{\partial\phi}{\partial\varphi_1}+\ldots
+\frac{\partial\phi}{\partial\varphi_n}\omega_n
$$
{\it on $\exp i(k,\,\varphi)$, apply the operation of averaging over}
$\mathbb{T}^n$
$$
\frac{1}{(2\pi)^n}\int\limits_{\mathbb{T}^n}(\,\cdot\,)\,
\partial \varphi_1\,\ldots, d\varphi_n,
$$
{\it and integrate by parts. In result we obtain the left part of relation
(20) up to the factor $-i$. The right part of (20) is obtained
similarly.}

Let us introduce the additional condition

\hangindent=37.5pt\hangafter=1 B) The Poincar\'{e} set $\mathbb{P}$ is
everywhere dense in any parallelepiped $K_i$.

The proposition 3 imply the following corollary

{\bf Corollary.} {\it
If conditions A and B are fulfilled, then the functions $\mathcal{F}_0$
and $\mathcal{F}_0$ are everywhere dependent.}

Hence we immediately obtain the following representation on any
parallelepiped $K_i$:
\begin{equation}
\mathcal{F}_0=F_i(\mathcal{H}_0),
\end{equation}
where $F_i$ is some continuously differentiable function.

Indeed, introducing the new variables
$H_1,\,\ldots,\,H_{n-1}$,\,$\mathcal{H}_0=\sum H_s$ instead of
$H_1,\,\ldots,\,H_n$ we can write down formula (11) in the another form:
$$
\mathcal{F}_0=f_i(H_1,\,\ldots,\,H_{n-1},\;
\mathcal{H}_0-H_1-\ldots-H_{n-1}).
$$
Since functions $\mathcal{F}_0$ and $\mathcal{H}_0$ are dependent, then
the right part of this equality actually does not depend on
$H_1,\,\ldots,\,H_{n-1}$.

\paragraph{6. Deduction of Gibbs distribution.}

Now let $\varepsilon$ tend to zero. At the limit we obtain the $n$
unjoined subsystems moving independently: the change of the initial data
$p^{(l)}$, $q^{(l)}$ for $l\ne s$ does not affect the dynamics of
subsystem with number $s$. In order to be consistent we shall also assume
at $\varepsilon=0$ that the subsystem with number $s$ being in some fixed
state $(p^{(s)},\,q^{(s)})\in M$ is a random event. The following natural
condition plays the main role in the process of deduction of Gibbs
distribution

C) These random events are independent.

If $\rho_s(p^{(s)},\,q^{(s)})$,\,$1\le s\le n$ is a density of probability
distribution of the subsystem with number~$s$, and
$$
\rho(P,\,Q,\,\varepsilon)=\rho_0(P,\,Q)+O(\varepsilon)
$$
is a density of probability distribution in the initial system with
Hamiltonian (2), then using condition C and the rules of multiplication
of probabilities of independent events we obtain as $\varepsilon\to0$ the
following equality:\vspace{-2mm}
\begin{equation}
\rho_0=\rho_1\,\ldots\,\rho_n.
\end{equation}

Equality (22) is also called the Gibbs hypothesis on the preservation of
thermodynamic equilibrium of subsystems at vanishing interaction ([1], see
also [13]). The sense of this term will be explained below.

Our main result is the following theorem

{\bf Theorem.}
{\it Suppose conditions A, B and C are fulfilled. Then}
\begin{equation}
\rho_s=c_s e^{-\frac{H_s}{k\tau}},\qquad c_s=const>0
\end{equation}
{\it for all} $1\le s\le n$.

In particular, according to (22),\vspace{-4mm}
$$
\rho_0=c_0 e^{-\frac{H_s}{k\tau}},\qquad c_0=c_1\ldots\,c_n.
$$
From (22) we see that all separate subsystems have the same distribution.
We compute the factors~$c_s$ using the normalizing condition
$$
\int\limits_M \rho_s\,d^np\:d^nq=1.
$$

{\it Proof of theorem.} First note that $\rho_s$ is a function of Hamiltonian
$H_s$ only, and it is continuously differentiable in all open
intervals\vspace{-3mm}
\begin{equation}
(a_1,\,a_2),\quad (a_2,\,a_3),\,\ldots,\quad (a_r,\,\infty).
\end{equation}
More exactly the number of such functions is equal to the number of the
connected components of level surface $\sum(h_s)$, when the energy $h_s$
changes in each of intervals (24). Some of these functions may coincide.

Indeed, for almost all $h_s\ge a_1$ the Hamiltonian system with
Hamiltonian $H_s$ has everywhere dense set of nondegenerate periodic
trajectories on energy surfaces $\sum(h_s)$. Otherwise condition A is not
fulfilled because of the identity of separate subsystems. Then, according
to Poincar\'{e} [11], the points of nondegenerate periodic trajectories
are stationary for the restriction of any first integral on $\sum(h_s)$.
The continuity imply that the first integrals of Hamiltonian system with
Hamiltonian function $H_s$ are constant on the connected components of
$\sum(h_s)$. At last we should note that $\rho_s$ is the first integral
and use the (simplified) arguments of section 4.

Now let us consider again equation (22) true on any parallelepiped $K_i$:
\begin{equation}
\rho_0(H_1+\ldots+H_n)=\rho_1(H_1)\,\ldots\,\rho_n(H_n).
\end{equation}
This functional equation is easily solved. We differentiate (25)
sequentially with respect to $H_1,\,\ldots,\,H_n$ and divide the result on
the product $\rho_1\,\ldots\,\rho_n$. In result we obtain the following
chain of equations
$$
\frac{\rho_1'}{\rho_1}=\ldots=\frac{\rho_n'}{\rho_n}=-\beta.
$$
Here $\beta$ is some constant independent on the number $s$. Hence
\begin{equation}
\rho_s=c_s e^{-\beta H_s},\qquad c_s=const.
\end{equation}
The dimension of constant $\beta$ is equal to the inverse energy
dimension. Usually one suppose that $\beta=(k\tau)^{-1}$, where
$\tau$ is the absolute temperature, and $k$ is the Boltzmann constant.

We should note that formula (26) may depend on the choice of multiindex
$i=(i_1,\,\ldots,\,i_n)$. More precisely, any connected component of the
set $\mathcal{K}_i\subset\mathcal{M}$ has its own set of the factors
$\beta$ and $c_s$ in (26).

However, we can easily show that the constant $\beta$ has an universal
character. Indeed, suppose the constants $\beta$ in formula (26) are
equal to the values $\beta_1,\,\ldots,\,\beta_n$ on some connected
component of domain $\mathcal{K}_i$ with some index $i$. Then, according
to (6.1), in this domain
\begin{equation}
\rho_0=c_0 e^{\sum\beta_s H_s}.
\end{equation}
If some $\beta_s$ are not equal to each other, then functions (27) and
$\mathcal{H}_0=\sum(H_s)$ are independent. But this statement contradicts
to the corollary of proposition 3.

This argument has the evident physical meaning: at thermodynamic
equilibrium all components of system have identical temperature.

The last remaining possibility is that constants $c_s$ in formula (26)
are different on the different intervals of values of energy (24). But in
reality this possibility is not realized because of the continuity
property of functions $\rho_s:M\to\mathbb{R}$.

The theorem is completely proved.

In conclusion of the work we shall make one important remark. The Gibbs
theory presented in [1] does not imply that the densities of probability
distributions $\rho_1,\,\ldots,\,\rho_n$ should be continuous functions on
$M$. We can consider more general situation and assume, for example, that
functions $\rho_s$ are continuously differentiable only on those domains
of phase space $M$, in which energy is contained between its neighboring
critical values
\begin{equation}
a_r<H_s<a_{r+1}\qquad (r=1,\,\ldots,\,p;\quad a_{p+1}=\infty).
\end{equation}
Naturalness of such assumption is already evident if we consider the
example of mathematical pendulum: the separatrices on phase cylinder
separate the domains with essentially different type of motion
(fluctuations and rotations).

Applying the developed above method we again obtain formula (23), but the
constants $c_s $ have different values in different domains (28).
Moreover, these constants may be different for different connected
components of domains (28). It is quite possible that such generalized
discontinuous Gibbs distribution could be useful for the study of concrete
thermodynamic systems.

Let us assume, for example, that the phase space $M$ has only two domains
$M_+$ and $M_-$ of form (28). In domains $M_\pm$ we have the following
densities of distributions
$$
c_\pm e^{-\frac{H}{k\tau}}.
$$
We calculate the constants $c_+$ and $c_-$ using the normalizing condition
$$
c_+\int\limits_{M_+}e^{-\frac{H}{k\tau}}d^np\,d^nq+
c_-\int\limits_{M_-}e^{-\frac{H}{k\tau}}d^np\,d^nq=1.
$$
If the difference $\Delta=c_+-c_-$ is given, then factors $c_\pm $ are
uniquely defined by this equality:
\begin{gather*}
c_+\int\limits_M e^{-\frac{H}{k\tau}}d^np\,d^nq=1+
\Delta\int\limits_{M_-}e^{-\frac{H}{k\tau}}d^np\,d^nq,\\
c_-\int\limits_M e^{-\frac{H}{k\tau}}d^np\,d^nq=1-
\Delta\int\limits_{M_+}e^{-\frac{H}{k\tau}}d^np\,d^nq.
\end{gather*}
Since $c_\pm>0$, then the right parts of these equalities are positive. It
happens with certainty if the jump $\Delta$ satisfies the following
inequality
$$
|\Delta|< \Biggl[\,\int\limits_M
e^{-\frac{H}{k\tau}}d^np\,d^nq\,\Biggr]^{-1}.
$$

Could we obtain the probability distribution with piecewise smooth
function of distribution using the Fowler-Darwin method?

The paper is prepared with the financial support of RFBR(99-01-0196) and
INTAS.

\begin{figure}[ht!]
$$
\includegraphics{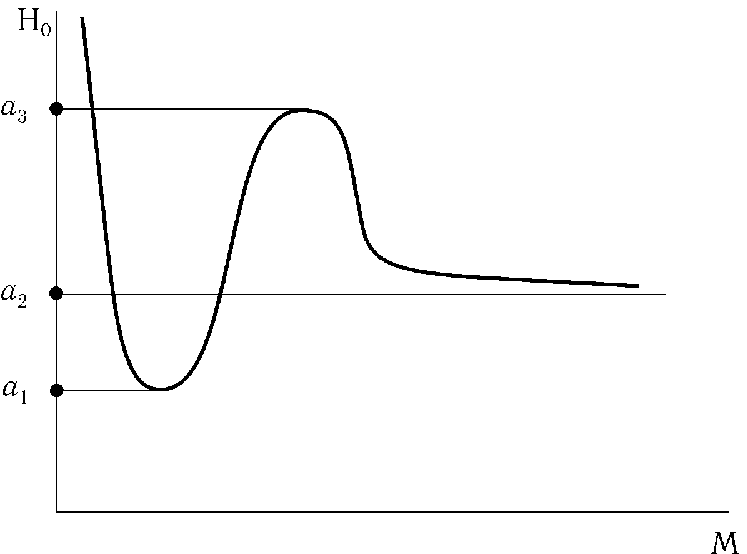}
$$
\caption{}
\end{figure}

\begin{figure}[ht!]
$$
\includegraphics{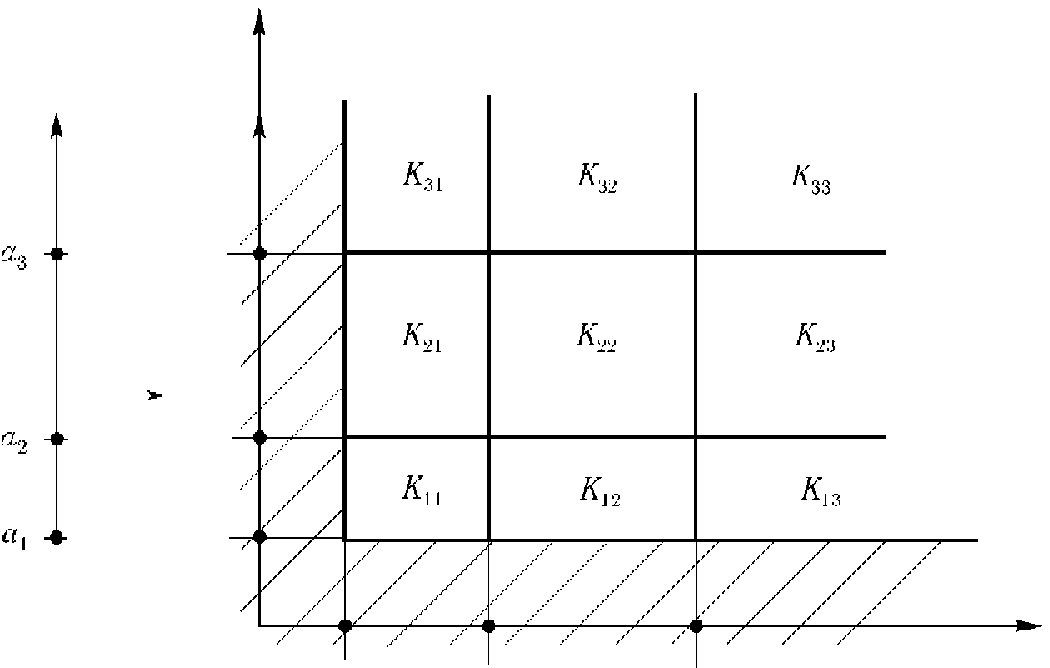}
$$
\caption{}
\end{figure}


\vspace{-7mm}


\begin{thebibliography}{00}

\bibitem{1}
J.\,W.\,Gibbs. {\it Elementary principles in statistical mechanics, developed
    with especial reference to the rational foundations of thermodynamics.}
    {N.\,Y.}{1902}

\bibitem{2}
L.\,Boltzmann. {\it Vorlesungen \"{u}ber Gastheorie.}
    {Leipzig: Ambrosius Barth}. {1912}

\bibitem{3}
D.\,Sz\'{a}sz. {\it Half Ball systems and the Lorentz Gas.}
    {Springer--Verlag}. {2000}

\bibitem{4}
R.\,H.\,Fowler, E.\,A.\,Guddenheim. {\it Statistical Thermodynamics.}
    {Cambridge: Univ. Press}. {1939}

\bibitem{5}
A.\,Ya.\,Hincin. {\it Mathematical justification of statistical mechanics.}
    {M.-L.: Gostechizdat}. {1943}

\bibitem{6}
F.\,A.\,Beresin. {\it Lectures on statistical physics.}
    {M.: MGU publ.} {1972}

\bibitem{7}
V.\,I.\,Arnold.{\it On instability of dynamical systems with many degrees of freedom.}
    Dokl. of USSR Acad. of Sciences, Vol. 156. No. {1}. {1964}, p.{9--12}

\bibitem{8}
A.\,J.\,Lichtenberg, M.\,A.\,Liebermann. {\it Regular and Stochastic Motion.}
    Springer--Verlag. 1983

\bibitem{9}
V.\,V.\,Kozlov, N.\,G.\,Moschevitin. {\it On diffusion in Hamiltonian systems.}
    Moscow Univ. Bull. Math. and mech. series. 1997, No. 5, p. 49--52

\bibitem{10}
V.\,V.\,Kozlov.{\it Canonical Gibbs distribution and thermodynamics of
    mechanical systems with a finite number of degrees of freedom.}
    Regular and Chaotic Dynamics, Vol. 4, No. {1}, 1999, p. 44--54

\bibitem{11}
H.\,Poincar\'{e}. {\it Les m\'{e}thode nouvelles de la m\'{e}canique
    celeste.} Vol. 1--3, Paris: Gauthier--Villars. 1892, 1893, 1899

\bibitem{12}
V.\,V.\,Kozlov.{\it Symmetries, Topology, and Resonances in Hamiltonian Mechanics.}
    Springer--Verlag. {1996}

\bibitem{13}
A.\,Sommerfeld. {\it Thermodinamik und Statistik.}
    Wissbaden, {1952}
\end{thebibliography}
\end{document}